\documentclass[conference]{IEEEtran}
\IEEEoverridecommandlockouts
\usepackage{cite}
\usepackage{amsmath,amssymb,amsfonts}
\usepackage{algorithmic}
\usepackage{graphicx}
\usepackage{textcomp}
\def\BibTeX{{\rm B\kern-.05em{\sc i\kern-.025em b}\kern-.08em
    T\kern-.1667em\lower.7ex\hbox{E}\kern-.125emX}}
\begin{document}

\title{Generating and Estimating Nonverbal Alphabets \\ for Situated and Multimodal Communications\\
\thanks{The work was partially supported by Ukraine-France Collaboration Project (Programme PHC DNIPRO) (http://www.campusfrance.org/fr/dnipro).}
}

\author{\IEEEauthorblockN{Serhii Hamotskyi*, Sergii Stirenko, Yuri Gordienko}
\IEEEauthorblockA{\textit{National Technical University of Ukraine} \\
\textit{"Igor Sikorsky Kyiv Polytechnic Institute"}\\
Kyiv, Ukraine \\
*shamotskyi@gmail.com}
\and
\IEEEauthorblockN{Anis Rojbi}
\IEEEauthorblockA{\textit{Laboratory CHArt (Human and Artificial Cognitions,} \\
\textit{University Paris 8,}\\
Paris, France \\
anis.rojbi@univ-paris8.fr}
}

\maketitle

\begin{abstract}
In this paper, we discuss the formalized approach for generating and estimating symbols (and alphabets), which can be communicated by the wide range of non-verbal means based on specific user requirements (medium, priorities, type of information that needs to be conveyed). The short characterization of basic terms and parameters of such symbols (and alphabets) with approaches to generate them are given. Then the framework, experimental setup, and some machine learning methods to estimate usefulness and effectiveness of the nonverbal alphabets and systems are presented. The previous results demonstrate that usage of multimodal data sources (like wearable accelerometer, heart monitor, muscle movements sensors, brain-computer interface) along with machine learning approaches can provide the deeper understanding of the usefulness and effectiveness of such alphabets and systems for nonverbal and situated communication. The symbols (and alphabets) generated and esrtimated by such methods may be useful in various applications: from synthetic languages and constructed scripts to multimodal nonverbal and situated interaction between people and artificial intelligence systems through Human-Computer Interfaces, such as mouse gestures, touchpads, body gestures, eye-tracking cameras, wearables, and brain-computing interfaces, especially in applications for elderly care and people with disabilities.
\end{abstract}


\section{Introduction}
The human behavior is usually situated and dependent on the environment and context. In the case of people with disabilities, the limited functional abilities put the additional constraints on their means and possibilities to communicate. Therefore, the modern technologies designed for processing human behavior and communication should be situated and address highly specific aspects of the contexts where they are used. In addition, some people with disabilities cannot principally communicate by verbal means, and the need for nonverbal methods and tools of human-human and human-machine interactions is growing in these cases. Recently, several attempts were applied for analysis and synthesis of human behavior with a purpose to develop the new approaches for situated and multimodal communications~\cite{rickheit2006situated}. The main aim of this short progress paper is to present the formalized approach for generating and estimating symbols, which can be communicated by the wide range of non-verbal means. The section \emph{2.Background} gives the very short outline of the state of the art. The section \emph{3.Generating a General-Purpose Nonverbal Alphabet} contains the short characterization of basic terms and parameters of such alphabets, and approach to generate them. The section \emph{4.Estimating a General-Purpose Nonverbal Alphabet} proposes the framework, experimental setup, and some machine learning methods to estimate usefulness and effectiveness of the nonverbal alphabets and systems. The section \emph{5.Discussion and future work} is dedicated to discussion of the results obtained and lessons learned.

\section{Background}
The problem of non-verbal communicative systems appeared from ancient times and nowadays it is evolved to the concrete challenges, especially in the context of human-machine and machine-human interactions~\cite{vinciarelli2015open}. In ancient times the technical side of the problem was related with limitations of data representation (gestures, mimics, body movements, dances, symbols, icons, letters, digits, etc.) on the available information carriers (from stone, clay, wood, metal, etc.). For example, the angular shapes of the runes were dictated by the necessity to carve them in wood or stone~\cite{williams1996origin}. But now explosive development of information and communication technologies allow us to widen the range of the non-verbal means of communication, which are already become de facto standards, for example, the proximity, gesture, haptic, and touch sensoric communication means in modern electronic carriers like mobiles, tablets, etc. The rapid increase of available mediums in the recent decades determined the need for many more tacit and informative means for representation of various non-verbal communicative signals (non-verbal alphabets) for very different use cases, such as controlling computers using touchpads, mouse gestures or eye tracking cameras. It is especially important for elderly care applications~\cite{gordienko2017augmented} on the basis of the newly available information and communication technologies with multimodal interaction through human-computer interfaces like wearable computing, augmented reality, brain-computer interfaces, etc~\cite{stirenko2017user}. 

\section{Generating a General-Purpose Nonverbal Alphabet}
As it is well-known nonverbal communication involves both conscious and unconscious processes of encoding and decoding. Encoding is related with generating information (such as facial expressions, gestures, body signs, etc.), and decoding is connected with the interpretation of the information obtained from various sensors from the initiator of the encoding information~\cite{craighead2004concise}. Many approaches for the manual creation of non-verbal alphabets have been used, but they are usually suboptimal, because encoding-decoding sometimes related with an inadequate cognitive load for people with various mentality. For example, the V sign (a hand gesture in which the index and middle fingers are raised and parted, while the other fingers are clenched) has different aтd sometimes opposite meanings, depending on the context and the type of presentation. After the Second World War it was used as a "V for Victory" sign, especially when Prime Minister Winston Churchill used it in his speeches~\cite{churchill1943photo}, but during the Vietnam War, in the 1960s, the "V sign" was widely adopted as a protest symbol of peace by some famous people like Yoko Ono~\cite{tuleja2012curious}. In a similar fashion, many systems do not use the possibilities given by the medium or context, electing to base themselves on already existing (familiar to the user, but suboptimal context-wise) symbols. A formalized framework capable of gathering requirements, generating symbols, grading them on a set of criteria and mapping them to meanings may be able to overcome many of those limitations.

\begin{figure}[h]
\centering
        \includegraphics[scale=0.75]{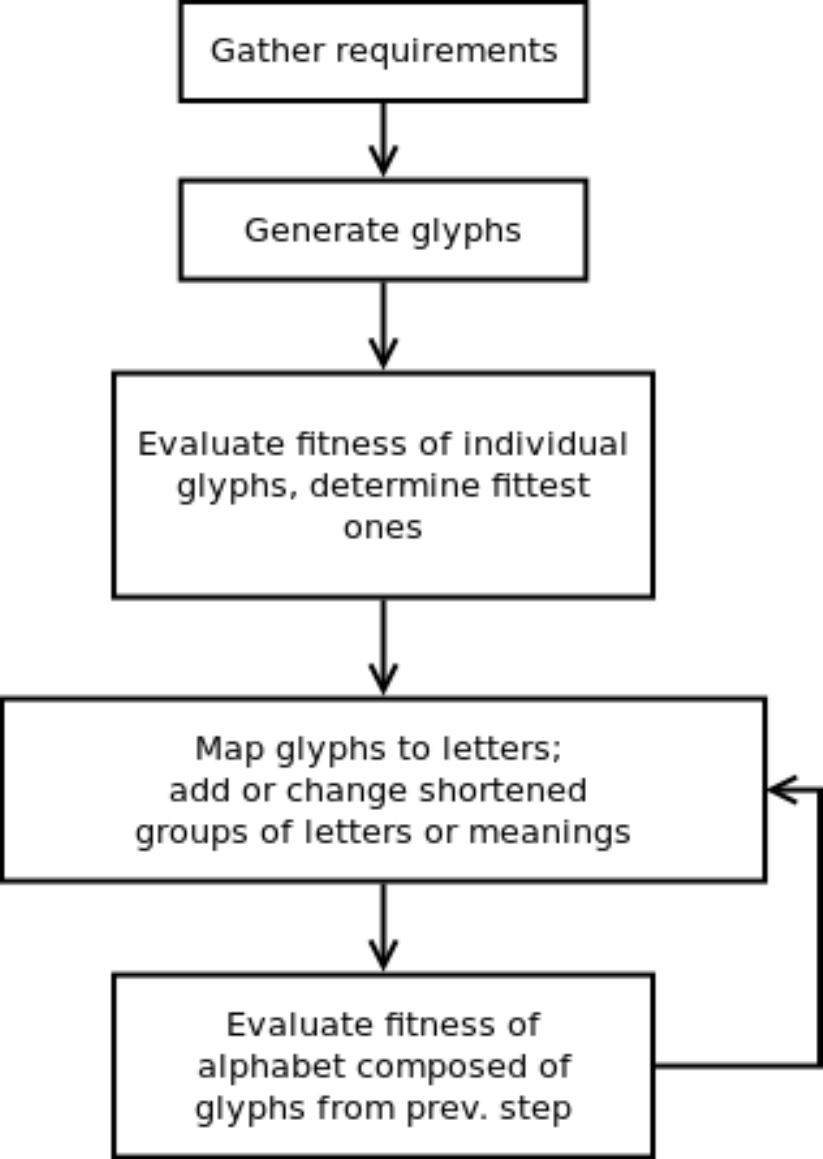}
        \caption{The basic workflow inside the proposed framework.}
        \label{fig_basic_workflow}
\end{figure}

The basic approach as proposed consists of several steps shown in Fig.~\ref{fig_basic_workflow}. In this paper, "glyph" is defined as unique mark/symbol in a given context and medium, and 2D symbols without varying width are used as its examples here and elsewhere~\cite{hamotskyi2017automatized}. Then "symbol" is defined as a glyph with some meaning attached to it, and nonverbal "phrase" is a sequence of such symbols. And "alphabet" is defined as a system of such symbols, including possible modifiers and conventions.

In the proposed framework, glyphs are generated and rated first, and meanings are assigned later with taking into account the situated context; only then the alphabet as a whole is rated. This two-step process design choice is based on performance reasons (mutating individual glyphs and their meanings at the same time is too complex for any reasonably-sized alphabet) and is meant as a starting point for further research and adaptation. 

The following desirable characteristics (and possible metrics) are assumed to be pertinent to almost any nonverbal alphabet, independently from the medium, dimensionality, and purpose. The vocabulary related to writing 2D symbols by a pen or stylus is used at the moment~\cite{hamotskyi2017automatized}, but this can be replaced with any other device, including body gestures, face gestures, eye movements, etc.
For everyday usage of the proposed nonverbal alphabet, comfort becomes a main concern and can be crucial for its success, i.e. whether it can be used or not used at all. This is especially important aspect for elders and people with disabilities. Usually, comfort is defined as "how easy and enjoyable is to use", but for precise estimation the quantitative criteria should be used. They might depend on the following factors:
\begin{itemize}
    \item mental effort: to recall the familiar/unfamiliar nonverbal glyph (ease of recalling the known image), to connect the glyph and situated context, to recognize the familiar symbol (ease of deciphering its meaning), to link several familiar symbols (phrase) (ease of decoding the meaning of the phrase),
    \item physical effort (fluency/flow of alphabet): to visualize/write/mark the individual glyph (for example, some strokes might be easier to write if someone is right-handed, or holds his pen in a certain way), to connect several nonverbal glyphs, etc.
\end{itemize}

\section{Estimating a General-Purpose Nonverbal Alphabet}
Quantifying the above mentioned mental and physical efforts is non-trivial task. Additionally, the subjective reactions to the same glyph might vary between people due to different demographic parameters (like age, mentality, cultural and/or language background, etc.) and especially in the view of the situated context (recall "V sign" example above). Our previous attempts of the hand-written digits recognition on the standard MNIST dataset~\cite{lecun1998mnist} demonstrated that this might be a promising area to study with the help of machine learning, especially by the novel deep learning approaches~\cite{kochura2017comparative,kochura2017comparativeperformance}. The conclusions derived like "symbols similar to X perform (correlate) poorly with demographic parameter Y" would be valuable for estimation the available nonverbal alphabets and creating the new alphabets. It should be noted that some machine learning attempts mentioned in the previous researches on the handwritten digits without any context~\cite{kochura2017comparative} and other researches on the text recognition in the context of Google Street View images~\cite{mishra2012top,neumann2012real} were not related with context and any demographic parameters at all. As to the demographic parameters, the following questions are of great importance: which kind of glyph/symbol/alphabet is more familiar to them, how much new information are they probably able to learn from nonverbal symbols, how many nonverbal symbols they are accustomed to use in everyday life, how many new nonverbal symbols they are able to retain, do they have any own/local nonverbal symbols. 
In the view of these considerations, several metrics like mental effort, physical effort, fluency, writing speed, ease of recognition, and universality were were proposed and considered in details elsewhere~\cite{hamotskyi2017automatized}. 

\begin{figure}[!h]
\centering
        \includegraphics[height=5cm]{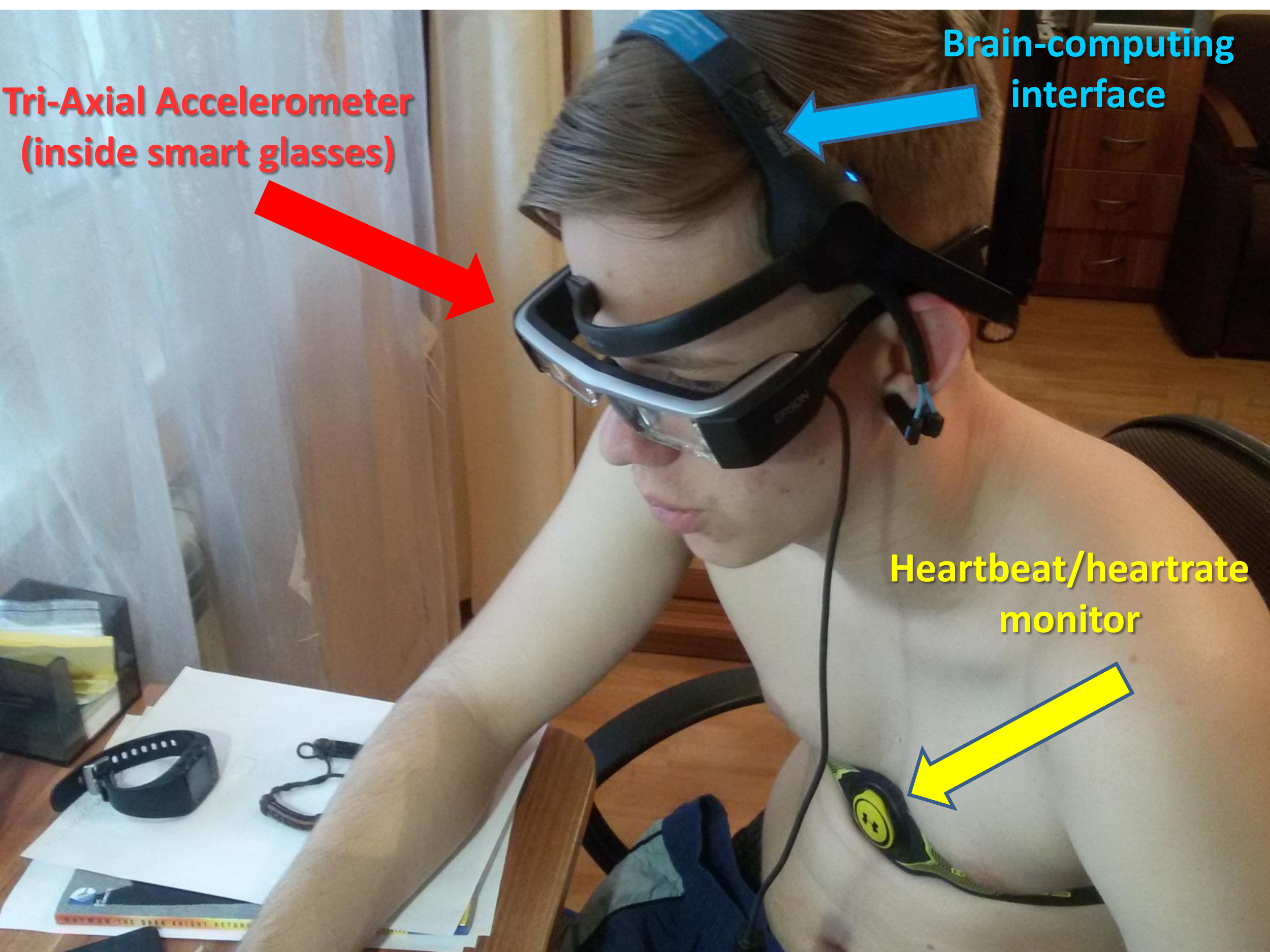} a) \includegraphics[height=5cm]{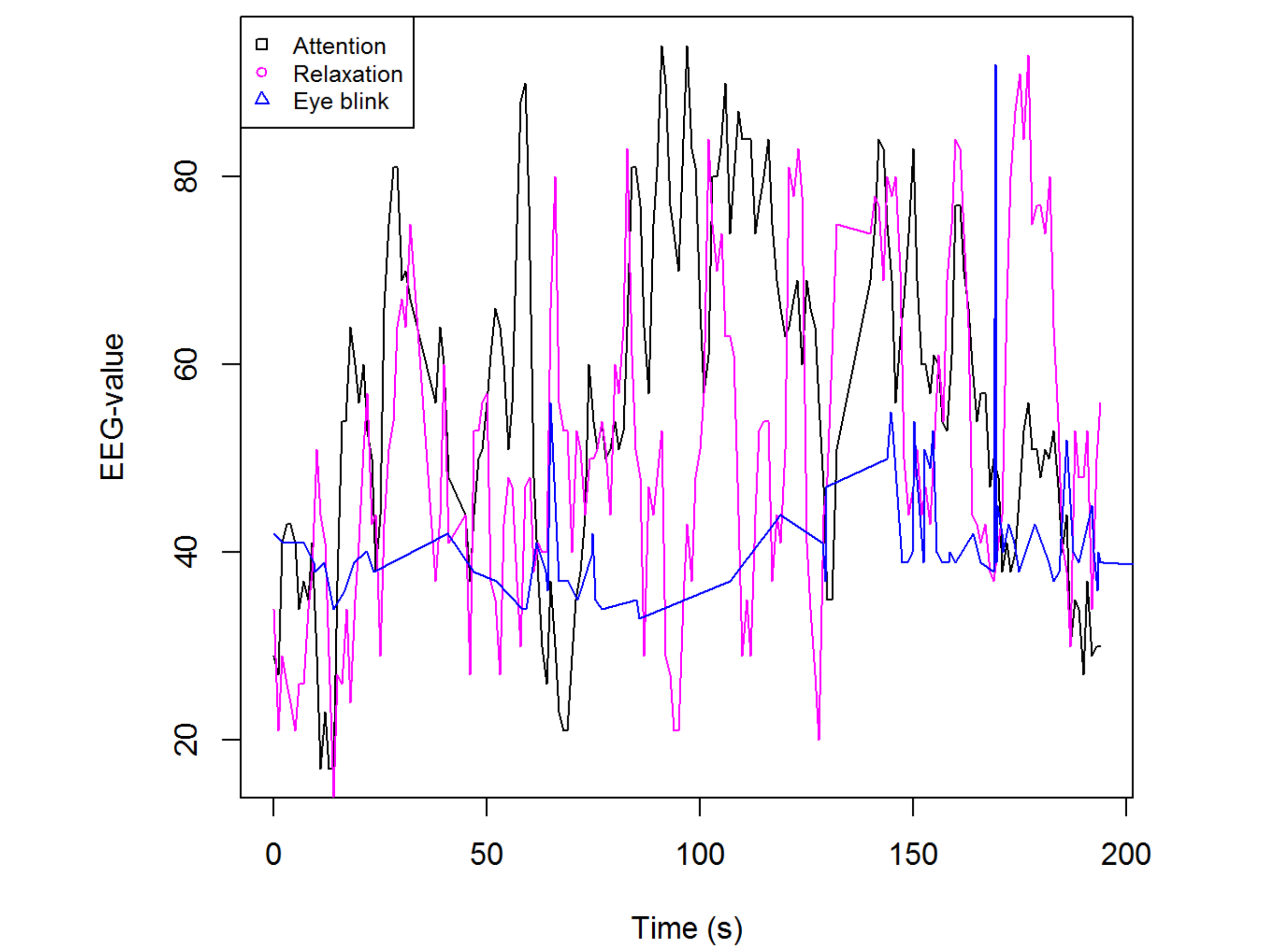} b)
        \caption{Experimental setup to measure mental and physical efforts (a) and example of BCI-data output (b) (see details in the text).}
        \label{fig_exp_setup}
\end{figure}

\begin{figure}[!h]
\centering
        \includegraphics[height=7cm]{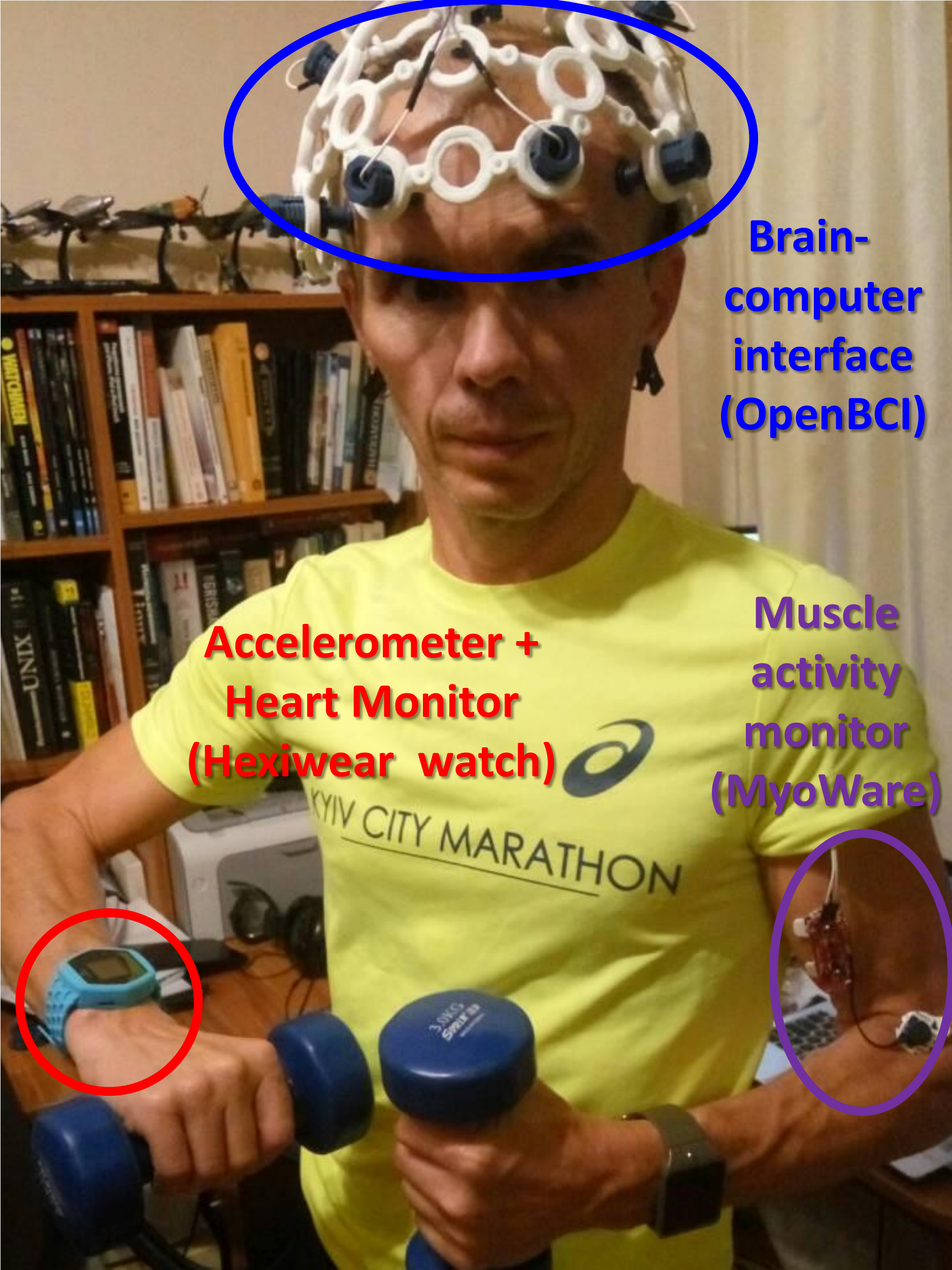}
        \caption{Advanced experimental setup to measure brain (brain-computer interface by OpenBCI~\cite{OpenBCI}), muscle (electromyography (EMG) sensors by MyoWare~\cite{Myoware}), and heart activities (heart monitor by Hexiwear~\cite{Hexiwear}) to measure mental and physical efforts (see details in the text).}
        \label{fig_exp_setup_02}
\end{figure}

Below, the approaches and metrics for their quantitative estimation for situated and multimodal nonverbal communications are proposed and discussed. Recently several approaches of fatigue estimation were proposed on the basis of multimodal human-machine interaction and machine learning methods~\cite{gordienko2017ccp}. In the context of measurements of the mental and physical efforts during nonverbal communication we used the experimental setup where the person under investigation performed several mental and physical actions and several sensoric data channels were used to measure his/her response (see Fig.~\ref{fig_exp_setup}a). The most promicing feedback as to the mental activity and efforts can be obtained by the advanced experimental setup with the more specific and accurate devices on the basis of multichannel brain-computer interface like OpenBCI, which is an open source brain-computer interface platform, created by Joel Murphy and Conor Russomanno~\cite{OpenBCI} (see Fig.~\ref{fig_exp_setup_02}). In the similar way the more specific information now is gathered by the locally situated muscle sensors (like Myoware~\cite{Myoware}) and heart rate monitors (breast heart monitor like UnderArmour 39 and wrist heart monitor like Hexiwear~\cite{Hexiwear}) (see Fig.~\ref{fig_exp_setup_02}).

The first acceleration channel included measurements of subtle tremors of head by accelerometer incorporated inside EPSON Moverio BT-200 smart glasses. The second heart activity channel was based on measurements of heart rate (beats of heart per minute) and heart beat (in seconds) by UnderArmour 39 heart monitor, where heart beat is the duration of the cardiac cycle (the reciprocal of heart rate, for example, the heart rate of 120 beats/minute corresponds to the heart beat of 0.5 seconds. The third brain-computer interface (BCI) channel collected data by electroencephalography (EEG), which is a noninvasive monitoring method to record electrical activity of the brain with the electrodes placed along the scalp. For this purpose MindWave Mobile BCI device by NeuroSky was used, which is the widely available and low cost (\$70-200) in comparison to professional EEG-devices~\cite{stirenko2017user}. It can measure activities of various frequencies (so-called Alpha, Beta, Gamma, Delta, Theta rhythms), that have the following interpretation: in terms of relaxation and concentration: Alpha rhythm (8-13 Hz) as an indication of physical relaxation and relative mental inactivity, Beta rhythm (13-35 Hz) as an indication of mental concentration. The typical example of these measurements is shown in Fig.~\ref{fig_exp_setup}b. 

\begin{figure}[!b]
\centering
        \includegraphics[scale=0.3]{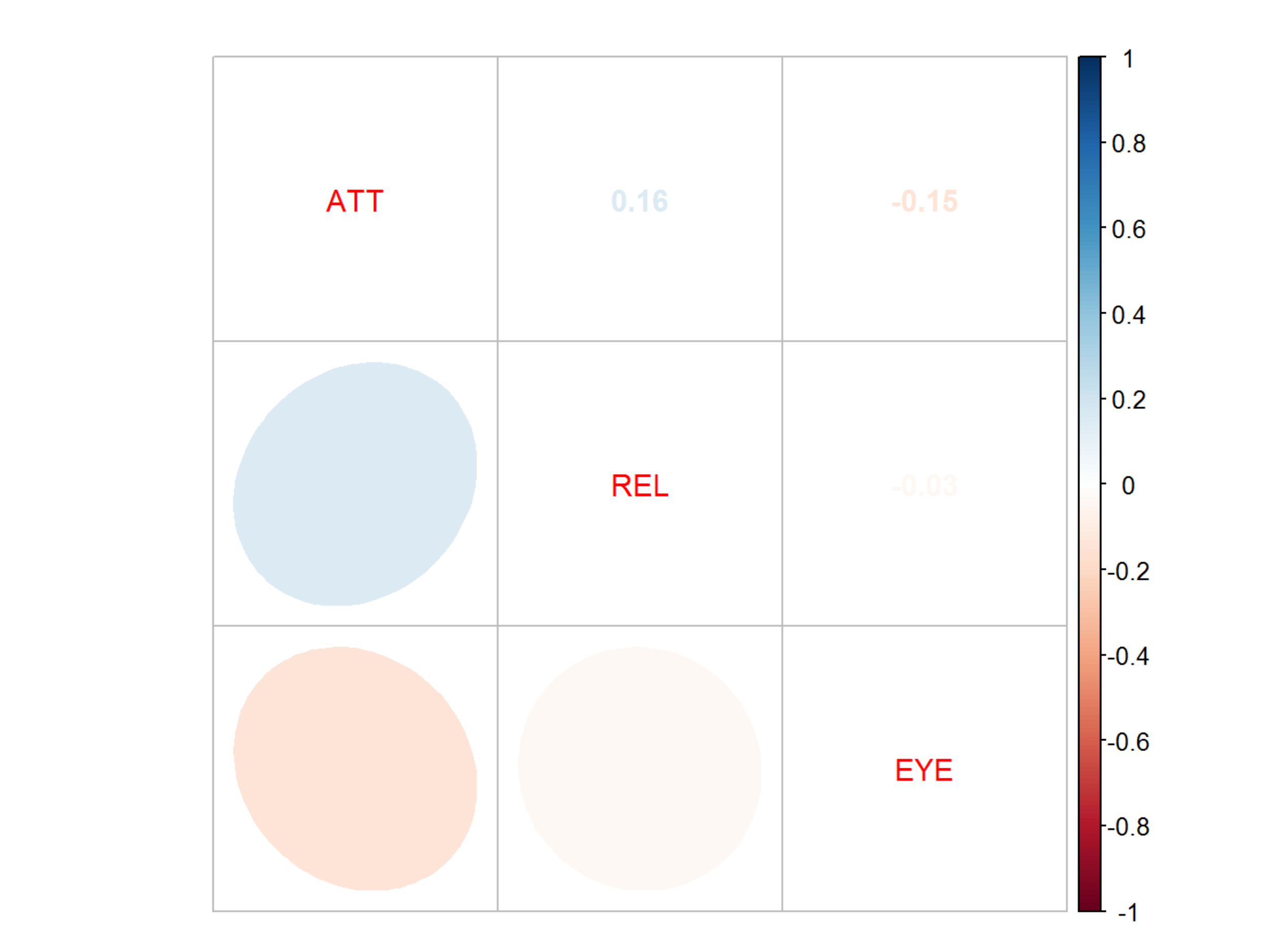}a) \includegraphics[scale=0.3]{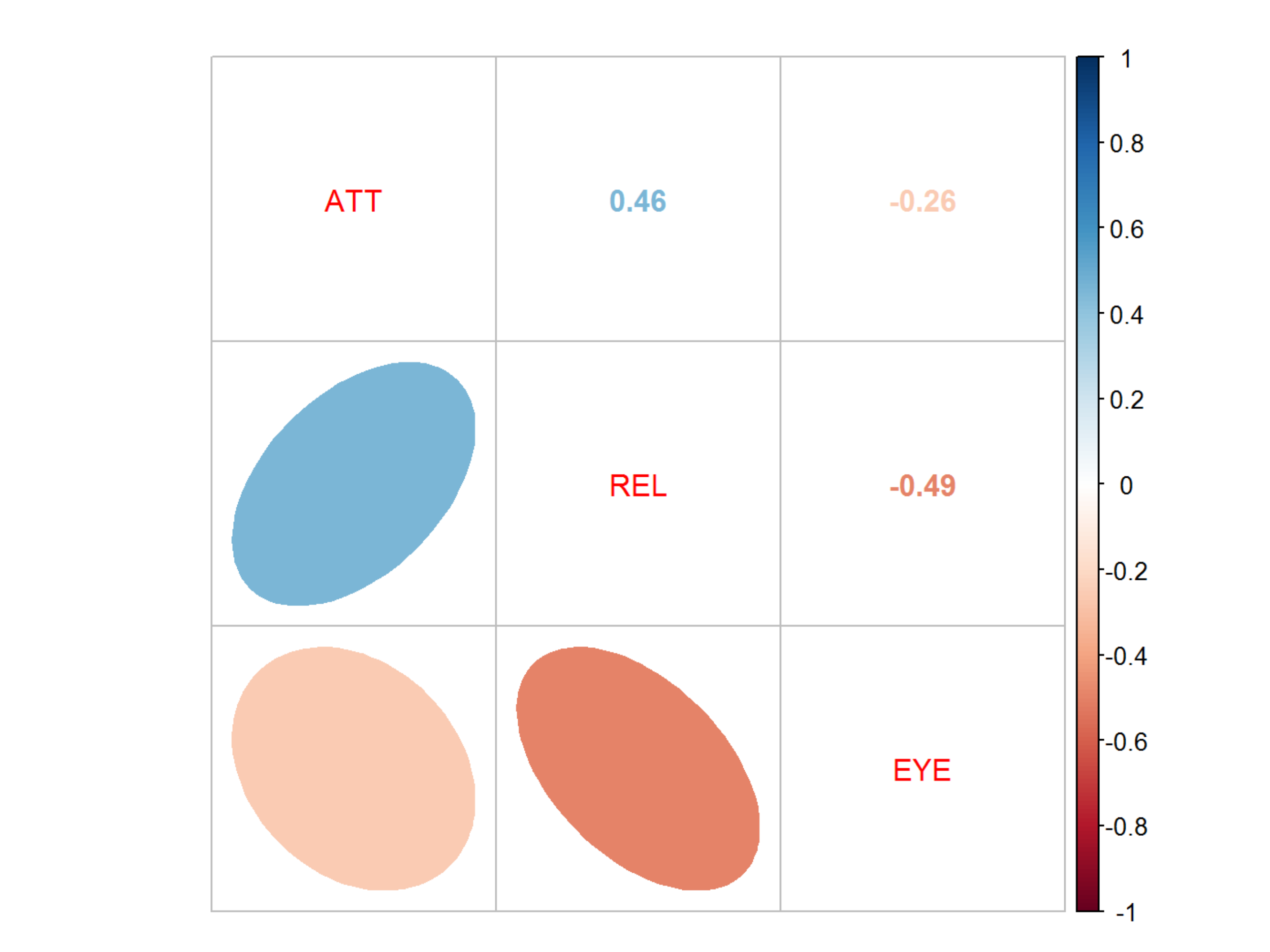}b)
        \caption{Correlation matrices for EEG activities measured as attention (ATT), relaxation (REL), and eye blink levels (EYE): (a) absolute values; (b) metrics as the moments of distributions of absolute values.}
        \label{fig_corr_plot}
\end{figure}

Several experiments were carried to measure response to various stimuli by these multimodal channels (acceleration, heart activity, and brain activity). The stimuli included mathematical operations, verbal and nonverbal communication in the various contexts and with different intensity. Their correlation analysis was performed for raw data values (Fig.~\ref{fig_corr_plot}a) and some parameters of their distributions (Fig.~\ref{fig_corr_plot}b).
This analysis allow us to make the previous conclusions that the parameters of the preprocessed data can provide the more targeted and sensitive characteristics (in comparison to raw data) of the response to the external stimuli, which was proved to be fruitful for other data mining purposes also~\cite{gordienko2015synergy}. The stimuli were provided mainly by three relatively different gesture activity: low (facial gestures only), medium (manipulations by fingers), and high (body signs).

\begin{figure}
\centering
        \includegraphics[scale=0.3]{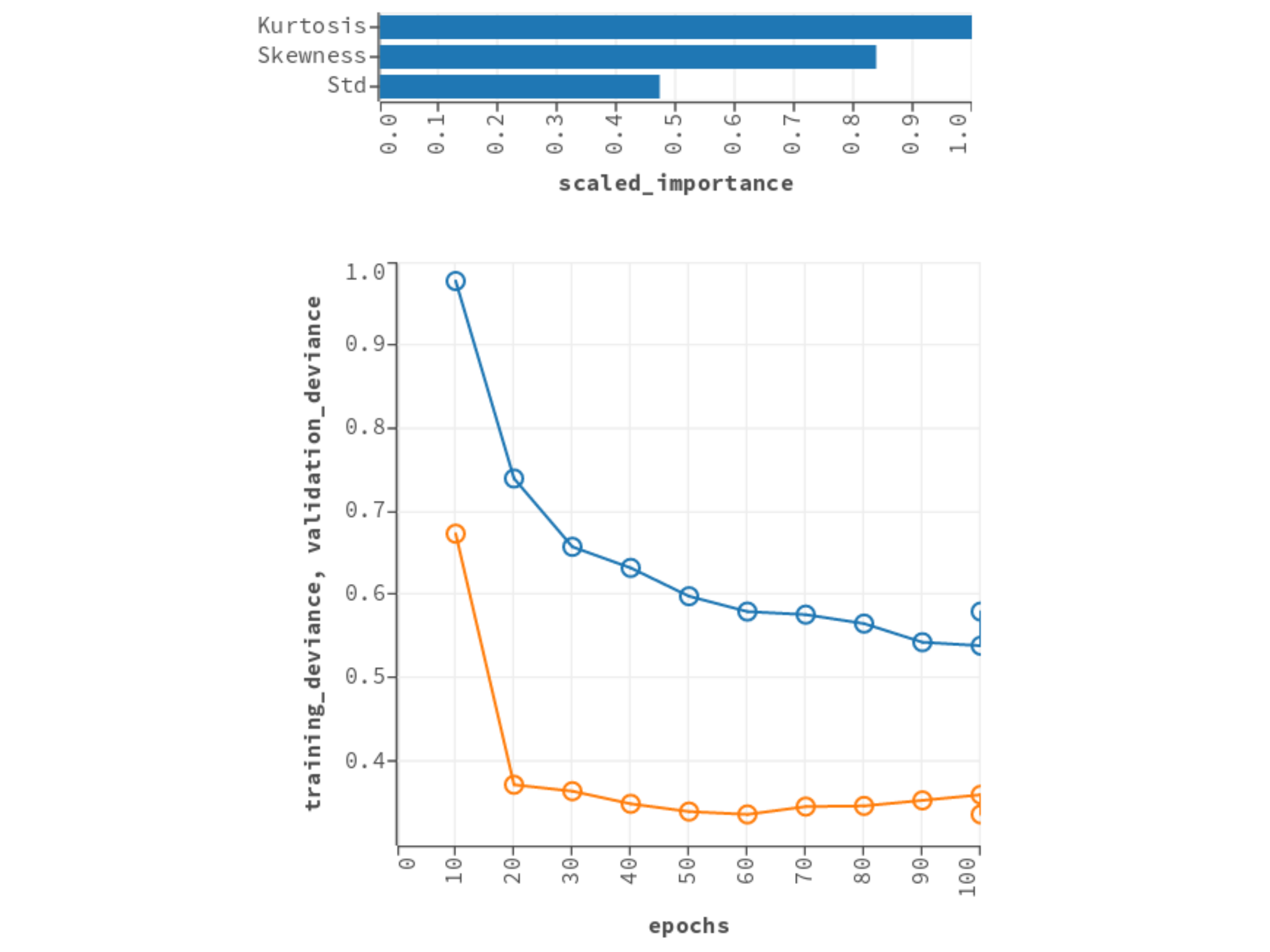}a) \includegraphics[scale=0.3]{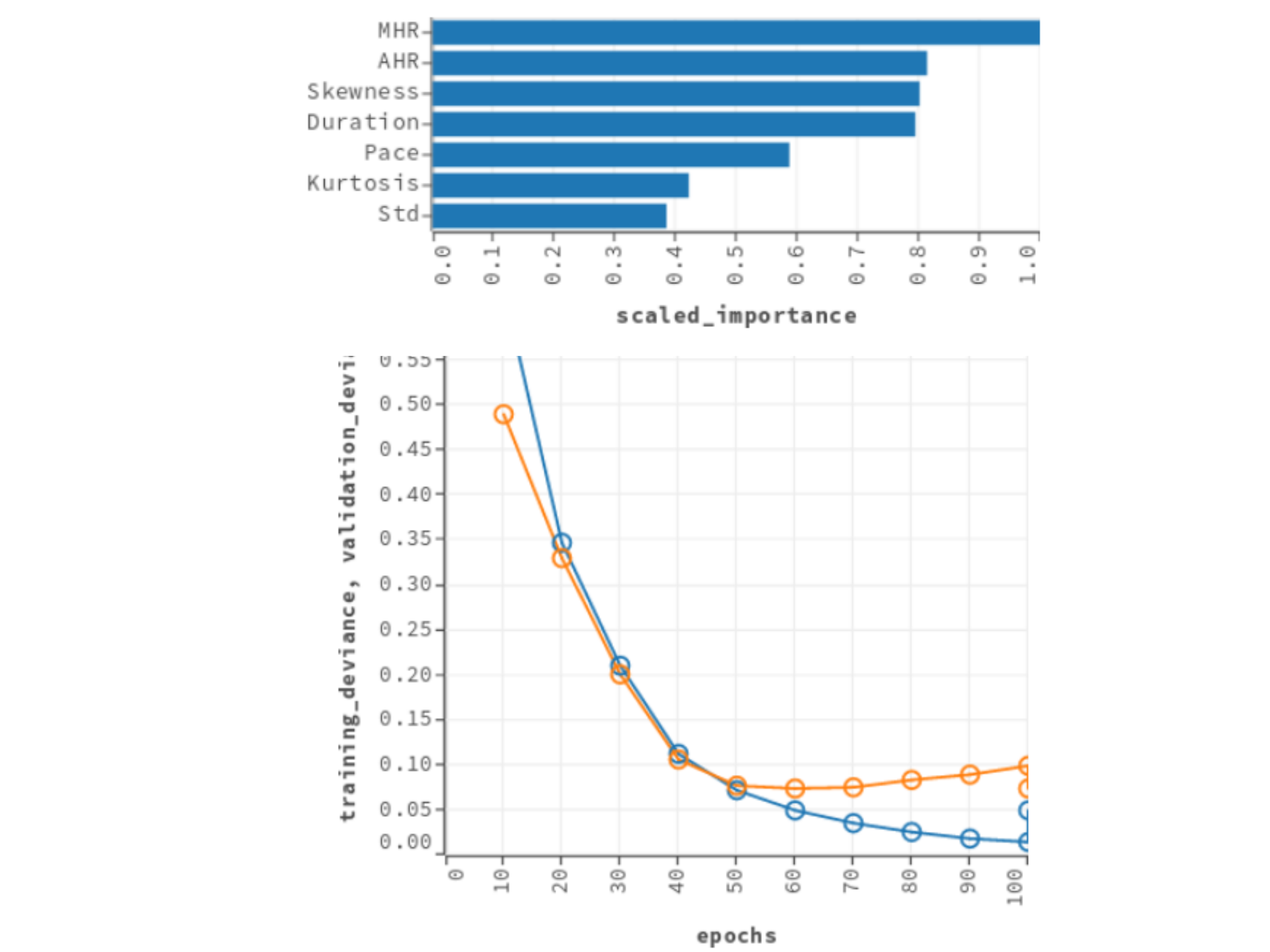}b)
        \caption{Machine learning results for the data obtained by multimodal channels like acceleration, heart activity, and brain activity (see details in the text).}
        \label{fig_ML}
\end{figure}

Application of machine learning methods to train and recognize intensity of the aforementioned nonverbal communication allowed us to analyze the mental and physical load on the persons under influence of various stimuli by these multimodal channels (acceleration, heart activity, and brain activity) including mathematical operations, verbal and nonverbal communication. The results obtained by machine learning methods,  actually, by deep learning neural networks~\cite{kochura2017ysf}, for the data obtained by multimodal channels (acceleration, heart activity, and brain activity) are shown in Fig.~\ref{fig_ML}. Fig.~\ref{fig_ML}a depicts the results for the short list of control parameters (only parameters of statistical distribution like standard deviation, skewness, kurtosis). And Fig.~\ref{fig_ML}b shows them for the long list of control parameters (parameters of statistical distribution + duration and pace of experiment, average (AHR) and maximal (MHR) heart rate). Here, the key parameters of the data collected and their relative influence on training the neural network (the upper diagrams), and the training (blue) and validation (yellow) rate with epochs of machine learning (the lower plots).     

\section{Discussion and future work}
At the moment the approaches demonstrated here are not specific enough to be considered seriously for applications, but they open several questions as to the possible ways for generation and estimation of the alphabets and systems for nonverbal communication. The main achievement is the multimodal data measured can be used as a training dataset for measuring and recognizing the intensity and physical load on the person involved in nonverbal communication by means of the machine learning approaches. The estimation of the mental load is the open question yet, because the 1-channel BCI device (MindWave Mobile by NeuroSky) is not precise and statistically reliable for the solid conclusions, but our previous analysis~\cite{stirenko2017user} shown that usage of the more powerful BCI devices (like multichannel OpenBCI~\cite{OpenBCI}) and EMG devices (like EMG-sensors by MyoWare~\cite{Myoware}) can be very promising in this context. The previous results demonstrate that usage of multimodal data sources (like wearable accelerometer, heart monitor, muscle movements monitor, brain-computer interface) along with machine learning approaches can provide the deeper understanding of the usefulness and effectiveness of such alphabets and systems for nonverbal and situated communication. In addition, one of the more interesting ways of further research would be creating algorithms for mapping glyphs to semantics with complex grammar categories or structures. Nevertheless, the presented ideas can be applied for the further investigation of nonverbal alphabets used for the real use cases of situated and multimodal communication. It is especially important to increase the range of ways that eye blinking, facial gestures, body gestures, and body movements exploit the non-linguistic context. In addition this approach open the new opportunities to exploit multimodal communication channels in different and personalized way, where each alphabet can be tailored and trained by machine learning to the functional abilities of users, for example, elders or people with disabilities.

\end{document}